\documentclass{article}
\usepackage{amssymb}
\usepackage{amsmath}

\input{tcilatex}

\begin{document}

\title{Replicator dynamical systems and their gradient and Hamiltonian
properties }
\author{A. Prykarpatsky\thanks{%
Institute of Applied Mathematics at the AGH University of Science and
Technology, Cracow, Polland, and Institute for Applied Problems of Mechanics
and Mathematics, National Academy of Sciences{} of Ukraine}, and V. V.
Gafiychuk\thanks{%
Institute for Applied Problems of Mechanics and Mathematics, National
Academy of Sciences{} of Ukraine, Naukova St,~3 b,~~Lviv~79601,~Ukraine}}
\date{\today}
\maketitle

\begin{abstract}
We consider the general properties of the replicator dynamical system from
the standpoint of its evolution and stability. Vector field analysis as well
as spectral properties of such system has been studied. Lyaponuv function
for investigation of system evolution has been proposed. The generalization
of the replicator dynamics for the case of multi-agent systems has been
introduced. We propose a new mathematical model to describe the multi-agent
interaction in complex system.

\noindent \textbf{keywords}: replicator dynamics, Hamiltonian systems,
gradient dynamical systems, multi-agent interaction, Lyapunov functions,
complex systems.
\end{abstract}

\section{Introduction}

Replicator equations were introduced by Fisher to capture Darwin's notion of
the survival of the fittest \cite{1a} and replicator dynamics is one of the
most important dynamic models arising in biology and ecology \cite{cam,mi},
evolutionary game theory \cite{hs} and economics \cite{yon,fr}, traffic
simulation systems and distributed computing \cite{kn} etc. It is derived
from the strategies that fare better than average and thriving is on the
cost of others at the expense of others (see e.g. \cite{hs}). This leads to
the fundamental problem of how complex multi-agent systems widely met in
nature can adapt to changes in the environment when there is no centralized
control in the system. For complex \textquotedblleft
living\textquotedblright\ system such problem has been considered in \cite%
{lg}. By term complex multi-agent systems under consideration we mean one
that comes into being, provides for itself, and develops pursuing its own
goals \cite{lg}. In the same time replicator dynamics arises if the agents
have to deal with conflicting goals and the behavior of such systems is
quite different from the adaptation problem considered in \cite{lg}. In this
article we develop a new dynamic model to describe the multi-agent
interaction in complex systems of interacting agents sharing common but
limited resources.

Based on this model we consider a population composed of $n\in Z_{+}$
distinct competing \textquotedblleft varieties\textquotedblright\ with
associated fitnesses $f_{i}(v)$, $i=\overline{1,n},$\ where $v\in \lbrack
0,1]$ is the vector of relative frequencies of the varieties $%
(v_{1},v_{2},...,v_{n})$. The evolution of relative frequencies is described
by the following equations:
\begin{equation}
dv_{i}/dt=v_{i}(f_{i}(v)-\left\langle f(v)\right\rangle ),~  \label{1}
\end{equation}%
where $i=\overline{1,n},$%
\begin{equation}
\left\langle f(v)\right\rangle =\sum\limits_{i=1}^{n}v_{i}f_{i}(v).
\label{2}
\end{equation}

The essence of (\ref{1}) and (\ref{2}) is simple: varieties with an
above-average fitness will expand, those with a below-average fitness will
contract.

Since $v_{i}\in \lbrack 0,1],i=\overline{1,n}$\ have to be nonnegative for
all time, the system (\ref{1}), (\ref{2}) is defined on the nonnegative
orthand

\begin{equation}
\mathbb{R}_{+}^{n}=\left\{ v\in \mathbb{R}^{n}:~v_{i}\geq 0\right\} .
\label{3}
\end{equation}
The replicator equation describes the relative share dynamics and thus holds
on the unit a n-1 dimensional space simplex
\begin{equation}
S_{n}=\left\{ v\in \mathbb{R}_{+}^{n}:\sum\limits_{i=1}^{n}v_{i}=1\right\} .
\label{s}
\end{equation}
\qquad

Let us write the Fisher's model in the following form

\begin{equation}
dv_{i}/dt=v_{i}\left( \underset{j=1}{\overset{n}{\sum }}a_{ij}v_{j}-\underset%
{j,k=1}{\overset{n}{\sum }}a_{jk}v_{j}v_{k}\right) ,  \label{p1a}
\end{equation}
where\ $a_{ij}\in \mathbb{R},$ $v_{i}\in \lbrack 0,1],$ $i,j=\overline{1,n},$
$n\in \mathbb{Z}_{+}$.

One can check the system (\ref{p1a}) can be written in such a matrix
commutative form:

\begin{equation}
dP/dt=\left[ \left[ D;P\right] ,P\right] .  \label{p4}
\end{equation}
Here by definition,

\begin{center}
\begin{equation}
P=\left\{ \left( v_{i}v_{j}\right) ^{\frac{1}{2}}:i,j=\overline{1,n}\right\},
\label{p4a}
\end{equation}

\begin{equation}
D=\frac{1}{2}diag\left\{ \underset{k=1}{\overset{n}{\sum }}a_{jk}v_{k}:j=%
\overline{1,n}\right\}.  \label{p5}
\end{equation}
\end{center}

It can be checked easily that the matrix $P\in $End$E^{n}$ is a projector in
$E^{n}$, that is $P^{2}=P$ for all $t\in \mathbb{R}$, what appears to be
very important for our further studying the structure of vector field (\ref%
{p4}) on the corresponding projector matrix manifold $\mathcal{P}$ \cite{5,7}%
. In particular, the expression (\ref{p4}) ensures that there exists the
time invariant vector subspace Im$P\in E^{n}$ in the Euclidian phase vector
space $E^{n},$ necessary for the replicating system under regard to be the
information data conserved.

\section{Vector field analysis}

In order to study the structure of the flow (\ref{p4}) on the projector
matrix manifold $\mathcal{P}\mathcal{\ni }P$ let us consider a functional $%
\Psi :\mathcal{P\rightarrow }\mathbb{R},$ where by definition the usual
variation

\begin{equation}
\delta \Psi (P):=\text{Sp}(D(P)\delta P),  \label{p7}
\end{equation}%
with $D\in \limfunc{End}E^{n},$ $\limfunc{Sp}:\limfunc{End}E^{n}\mathcal{%
\rightarrow }\mathbb{R}^{1}$ being the standard matrix trace. Taking into
account the natural metrics on $\mathcal{P}$, \ we \ consider the projection
of the usual gradient vector field $\nabla \Psi $ to the tangent space $T(%
\mathcal{P})$ under the following conditions:

\begin{equation}
\varphi (X;P):=\limfunc{Sp}(P^{2}-P,X)=0,\ \ \ \ \ \ \ \limfunc{Sp}(\nabla
\varphi ,\nabla _{\varphi }\Psi )\mid _{\mathcal{P}}=0,  \label{p8}
\end{equation}%
holding on $\mathcal{P}$ for all $X\in \limfunc{End}E^{n}.\;$The first
condition is evidently equivalent to $P^{2}-P=0,$ that is $P\in \mathcal{P}$
. Thereby we can formulate such a lemma.

\textbf{Lemma 1}. The functional gradient $\nabla _{\varphi }\Psi (P),$ $%
P\in \mathcal{P}$, at condition (\ref{p8}) has the following commutator
representation:

\begin{center}
\begin{equation}
\nabla _{\varphi }\Psi (P)=\left[ \left[ D,P\right] ,P\right] .  \label{ad}
\end{equation}
\end{center}

Proof. Consider the projection of the usual gradient $\nabla \Psi (P)$ to
the tangent space $T(\mathcal{P})$ of the manifold $\mathcal{P~}$having
assumed that $P\in \limfunc{End}E^{n}$:

\begin{equation}
\nabla _{\varphi }\Psi (P)=\nabla \Psi (P)-\nabla _{\varphi }(\Lambda ,P),
\label{p10}
\end{equation}%
where $\Lambda \in \limfunc{End}E^{n}$ is some unknown matrix. Taking into
account the conditions (\ref{p8}), we find

\begin{eqnarray}
\nabla _{\varphi }\Psi (P) &=&D-\Lambda -P(D-\Lambda )-(D-\Lambda )P+PD+DP
\label{p11} \\
&=&PD+DP+2P\Lambda P.  \label{p12a}
\end{eqnarray}
where\ we made use the conditions
\begin{equation*}
\nabla _{\varphi }\Psi (P)=D-\Lambda +P\Lambda +\Lambda P)
\end{equation*}
and
\begin{equation*}
P(D-\Lambda )+(D-\Lambda )P+2P\Lambda P=D-\Lambda .
\end{equation*}
Now one can see from (\ref{p12a}) and the second condition in (\ref{p10}),
that

\begin{equation}
P\Lambda P=-PDP  \label{p13}
\end{equation}%
for all $P\in \mathcal{P}$, giving rise to the final result

\begin{equation}
\nabla _{\varphi }\Psi (P)=PD+DP-2PDP,  \label{p14}
\end{equation}%
coinciding exactly with commutator (\ref{ad}).

It should be noted that the manifold $\mathcal{P}$ is also a symplectic
Grassmann manifold (\cite{5,7}), whose canonical symplectic structure is
given by the expression:

\begin{equation}
\omega ^{(2)}(P):=\limfunc{Sp}(PdP\wedge dPP),  \label{p15}
\end{equation}%
where $d\omega ^{(2)}(P)=0$ \ for all $P\in \mathcal{P}$, and the
differential form (\ref{p15}) is non-degenerate \cite{5,8} upon the tangent
space $T(\mathcal{P}).$

Let us assume that $\xi :\mathcal{P\rightarrow }\mathbb{R}$ is an arbitrary
smooth function on $\mathcal{P}$ . Than Hamiltonian vector field $X_{\xi }:%
\mathcal{P\rightarrow }T(\mathcal{P})$ on $\mathcal{P}$ generated by this
function relative to the symplectic structure (\ref{p15}) is given as
follows:

\begin{equation}
X_{\xi }=[[D_{\xi },P],P],  \label{p16}
\end{equation}%
where $D_{\xi }\in \limfunc{End}E^{n}$ is a certain matrix. The vector field
$X_{\xi }:\mathcal{P\rightarrow }T(\mathcal{P})$ generates on $\mathcal{P}$
the flow

\begin{equation}
dP/dt=X_{\xi }(P)  \label{p17}
\end{equation}
being defined globally for all $t\in \mathcal{\ }\mathbb{R}$. This flow by
construction is evidently compatible with the condition $P^{2}=P$. This
means in particular that

\begin{equation}
-X_{\xi }+PX_{\xi }+X_{\xi }P=0.  \label{p18}
\end{equation}
Thus, we stated that dynamical system (\ref{p4})being considered on the
Grassmann manifold $\mathcal{P~}$is Hamiltonian what makes it possible to
formulate the following statement.

\textbf{Statement 1.} A gradient vector field of the form (\ref{p16}) on
Grassmann manifold \ $\mathcal{P}$ is Hamiltonian with respect to the
canonical symplectic structure (\ref{p15}) and certain Hamiltonian function $%
\xi :\mathcal{P\rightarrow }\mathbb{R}$, satisfying

\begin{equation*}
\nabla \xi (P)=[D_{\xi },P],\ \ D_{\xi }=D,
\end{equation*}
where simultaneously $\nabla _{\varphi }\Psi (P)=X_{\xi }(P)$ for all $P\in
\mathcal{P}$.

Consider now the (n-1)-dimensional Riemannian space

\begin{equation*}
M_{g}^{n-1}=\left\{ v_{i}\in \mathbb{R}_{+}:i=\overline{1,n},\text{ }%
\underset{i=1}{\overset{n}{\sum }}v_{i}=1\right\}
\end{equation*}
with the metrics

\begin{equation*}
ds^{2}(v):=d^{2}\Psi \mid _{\mathcal{P}}(v)=\underset{i,j=1}{\overset{n}{%
\sum }}g_{ij}(v)dv_{i}dv_{j}\mid _{\mathcal{P}},
\end{equation*}
where

\begin{equation*}
g_{ij}(v)=\frac{\partial ^{2}\Psi (v)}{\partial v_{i}\partial v_{j}},\ i,j=%
\overline{1,n},\ \underset{i=1}{\overset{n}{\sum }}v_{i}=1.
\end{equation*}
Subject to the metrics on $M_{g}^{n-1}$ we can calculate the gradient $%
\nabla _{\varphi }\Psi $ of the function $\Psi :\mathcal{P\rightarrow }%
\mathbb{R}$ and set on $M_{\varphi }^{n-1}$ the gradient vector field

\begin{equation}
dv/dt=\nabla _{\psi }\Psi (v),  \label{p19}
\end{equation}
\ where $v\mathcal{\in }M_{\psi }^{n-1}$, or $\underset{i=1}{\overset{n}{%
\sum }}v_{i}=1$ is satisfied. Having calculated (\ref{p19}), we can
formulate the following statement.

\textbf{Statement 2.} The gradient vector fields $\nabla _{\varphi }\Psi $
on $\mathcal{P}$ and $\nabla _{g}\Psi $ on $M_{g}^{n-1}$ are equivalent or
in another words vector fields

\begin{equation}
dv/dt=\nabla _{g}\Psi (v)  \label{p20}
\end{equation}
and
\begin{equation}
dP(v)/dt=\left[ \left[ D(v),P(v)\right] ,P(v)\right]  \label{p20a}
\end{equation}
\ generates the same flow on $M_{g}^{n-1}$.

As a result from the Hamiltonian property of the vector field $\nabla
_{\varphi }\Psi $ on the Grassmann manifold $\mathcal{P}$ we get a new
statement.

\textbf{Statement 3}. The gradient vector field $\nabla _{g}\Psi $ (\ref{p19}%
) on the metric space $M_{g}^{n-1}$ is Hamiltonian subject to the
non-degenerate symplectic structure

\begin{equation}
\omega _{g}^{(2)}(v):=\omega ^{(2)}(P)\mid _{M_{g}^{n-1}}  \label{p21}
\end{equation}
\ for all $v\mathcal{\in }M_{g}^{2m}$ with the Hamiltonian function $\xi
_{\psi }:M_{g}^{2m}\mathcal{\rightarrow \mathbb{R}}$, where $\xi _{\psi
}:=\xi \mid _{M_{g}^{n-1}},$ $\xi :\mathcal{P\rightarrow }\mathbb{R}$ is the
Hamiltonian function of the vector field $X_{\xi }$ (\ref{p16}) on $\mathcal{%
P}$ . Otherwise if $n\in Z_{+}$ is arbitrary \ our two flows (\ref{p20}) and
(\ref{p20a}) are on $\mathcal{P}$ only Poissonian.

\section{Spectral properties}

Consider the eigenvalue problem for a matrix $P\in \mathcal{P}$, depending
on evolution parameter $t\in \mathbb{R}$:

\begin{equation}
P(t)f=\lambda f,  \label{p22}
\end{equation}%
where $f\in \mathbb{R}^{n}$ is an eigenfunction, $\lambda \in \mathbb{R}$ is
a real eigenvalue $P^{\ast }=P$, i.e. matrix $P\in \mathcal{P}$ is
symmetric. It is seen from expression (\ref{p20}) that $\limfunc{spec}%
P(t)=\left\{ 0,1\right\} $for all $t\in \mathbb{R}$. Moreover, taking into
account the invariance of $\limfunc{Sp}P=1$ we can conclude that \ only one
eigenvalue of the matrix $P(t),$ $t\in \mathbb{R}$, is equal to 1, all
others being equal to zero. So, we can formulate the next lemma.

\textbf{Lemma 2}. The image $\func{Im}P\subset E^{n}$ of the matrix $P(t)\in
\mathcal{P}$ for all $t\in \mathcal{\ }\mathbb{R}$ is k-dimensional $%
k=rankP,\ $and the kernel $\ker P\subset E^{n}$ is $(n-k)$-dimensional,
where $k\in \mathbb{Z}_{+}$ is constant, not depending on $t\in \mathcal{\ }%
\mathbb{R}$ . As a consequence of the lemma we establish that at $k=1$ there
exists a unique vector $f_{0}\in E^{n}/(\ker P)$ for which

\begin{center}
\begin{equation}
Pf_{0}=f_{0},\ \ f_{0}\simeq E^{n}/(\ker P).  \label{p23}
\end{equation}
\end{center}

Due to the statement above for projector $P:E^{n}\mathcal{\rightarrow }E^{n}$
we can write down the following invariant in time expansion in the direct
sum of mutually orthogonal subspaces:

\begin{center}
$E^{n}=\ker P\oplus \func{Im}P.$
\end{center}

Take now $f_{0}\in E^{n}$ satisfying the condition (\ref{p23}). Than in
accordance with (\ref{p20}) the next lemma holds.

\textbf{Lemma 3.} The vector $f_{0}\in E^{n}$ satisfies the following
evolution equation:

\begin{equation}
df_{0}/dt=\left[ D(v),P(v)\right] f_{0}+C_{0}(t)f_{0},  \label{p25}
\end{equation}%
where $C_{0}:\mathbb{R}\mathcal{\rightarrow }\mathbb{R}$ is a certain
function depending on the choice of the vector $f_{0}\in \func{Im}P$. At
some value of the vector $f_{0}\in \func{Im}P$ we can evidently ensure the
condition $C_{0}\equiv 0$ for all $t\in \mathcal{\ }\mathbb{R}^{n}$.
Moreover one easily observes that for the matrix $P(t)\in \mathcal{P}$ \ one
has \cite{7} \ the representation $P(t)=f_{0}\otimes f_{0}$, $\left\langle
f_{0},f_{0}\right\rangle =1$ , giving rise to the system (\ref{p1a}) if $%
f_{0}:=\left\{ \pm \sqrt{v_{j}}\in \mathbb{R}_{+}:j=\overline{1,n}\right\}
\in E^{n}$.

\section{Lyapunov Function}

Let us consider gradient vector fields $\nabla _{\varphi }\Psi $ on $%
\mathcal{P}$ and $\nabla _{g}\Psi $ on $M_{g}^{n-1}$. It is easy to state
that the function $\Psi :\mathcal{P\rightarrow }\mathbb{R}$ given by (\ref%
{p7}) and equal on $M_{g}^{n-1}$following expression
\begin{equation}
\Psi (v)=\underset{i,j=1}{\overset{n}{\frac{1}{4}\sum }}a_{ij}v_{i}v_{j}-c%
\underset{i=1}{\overset{n}{\sum }}v_{i}  \label{p29}
\end{equation}
\ under the condition $\ c\in \mathbb{R}_{+},$\ $\underset{i=1}{\overset{n}{%
\sum }}v_{i}=1$ being at the same time a Lyapunov function for the vector
fields $\nabla _{\varphi }\Psi $ on $\mathcal{P}$ and $\nabla _{g}\Psi $ on $%
M_{g}^{n-1}$. Indeed:

\begin{eqnarray}
\frac{d\Psi }{dt} &=&\langle \nabla _{g}\Psi ,\frac{dv}{dt}\rangle
_{T(M_{g}^{n-1})}  \label{p27} \\
&=&\langle \nabla _{g}\Psi ,\nabla _{g}\Psi \rangle _{T(M_{g}^{n-1})}\geq 0,
\notag \\
\frac{d\Psi }{dt} &=&\limfunc{Sp}(DdP)/dt=\limfunc{Sp}(D\frac{dP}{dt})
\label{p28} \\
&=&\limfunc{Sp}(D,\left[ \left[ D,P\right] ,P\right] )=-\limfunc{Sp}(\left[
P,D\right] ,\left[ P,D\right] )  \notag \\
&=&\limfunc{Sp}(\left[ P,D\right] ,\left[ P,D\right] ^{\ast })\geq 0  \notag
\end{eqnarray}
for all $t\in \mathbb{R}$ , where $\langle .,.\rangle _{T(M_{g}^{n-1})}$ is
the scalar product on $T(M_{g}^{n-1})$ obtained via the reduction of the
scalar product $\langle .,.\rangle $ on $\mathbb{R}^{n}$upon $T(M_{g}^{n-1})$
under the constraint \ $\underset{i=1}{\overset{n}{\sum }}v_{i}=1$.

\section{Multi-agent replicator system}

In the above we have considered the case when $rankP(t)=1,$ $t\in \mathbb{R}$%
. It is naturally to study now the case when $n\geq rankP(t)=k>1$ being
evidently constant for all $t\in \mathbb{R}$ too. This means therefore that
there exists some orthonormal vectors $f_{\alpha }\in \func{Im}P,$ $\alpha =%
\overline{1,k},$ that
\begin{equation}
P(t)=\sum\limits_{\alpha =1}^{k}f_{\alpha }(t)\otimes f_{\alpha }(t)
\label{p31}
\end{equation}%
for all $t\in \mathbb{R}$. Put now $f_{\alpha }:=\left\{ \pm \sqrt{%
v_{i}^{(\alpha )}}\in \mathbb{R}:i=\overline{1,n}\right\} \in ImP,$ $%
\sum\limits_{i=1}^{n}v_{i}^{_{(\alpha )}}=1$ for all $\alpha =\overline{1,k}%
. $ The necessary orthonormality condition $<f_{\alpha },f_{\beta }>=\delta
_{\alpha \beta },$ $\alpha ,\beta =\overline{1,k},$ can be automatically
satisfied if one put $f_{\alpha ,i}:=(\exp h)_{i,\alpha },~i=\overline{1,n},$
for a general skew-symmetric matrix $h=-h^{\ast }$ in the Euclidian vector
space $E^{n}.$ Then, evidently, all signs at vector components $\pm \sqrt{%
v_{i}^{\alpha }}=f_{\alpha ,i},i=\overline{1,n}$ will be defined exactly for
each $\alpha =\overline{1,k}.$ As a simple consequence of the above
representation for stable vector $f_{\alpha }\in E^{n},\alpha =\overline{1,k}
$ one derives that in the multiagent case one can not write down the
resulting replicator dynamics equations in the terms of positive
concentration frequencies. The corresponding gradient flow on $\mathcal{P}$
then takes modified commutator form with the Lyapunov type function
variation $\delta \Psi :\mathcal{P\rightarrow }\mathbb{R}$ as follows:
\begin{equation}
\delta \Psi =\sum\limits_{\alpha =1}^{k}Sp(D_{\alpha }\delta P^{(\alpha )}),
\label{p32}
\end{equation}%
where by definition $D_{\alpha }^{\ast }=D_{\alpha }\in endE^{n},$ and $%
P^{(\alpha )}=f_{\alpha }(t)\otimes f_{\alpha }(t),$ $P^{(\alpha )}P^{(\beta
)}=P^{(\alpha )}\delta _{\alpha \beta },$ $\alpha ,\beta =\overline{1,k},$ $%
\sum\limits_{\alpha =1}^{k}P^{(\alpha )}=P.$ The resulting flow on $\mathcal{%
P}$ is given as

\begin{equation}
dP/dt=\nabla _{\varphi }\Psi (P),  \label{p33}
\end{equation}%
where the gradient $\nabla _{\varphi }:\mathcal{D(P)\rightarrow }T\mathcal{%
(P)}$ is calculated taking into account the set of natural constraints:

\begin{equation}
\varphi _{_{\alpha ,\beta }}(X;P):=Sp\left[ X_{\alpha ,\beta }(P_{\alpha
}P_{\beta }-\delta _{\alpha ,\beta }P_{\beta })\right] =0  \label{p34}
\end{equation}%
for every $\alpha \leqslant \beta =\overline{1,k.}$ Assume now that $%
k=rankP(t)$ for all $t\in \mathbb{R}$ \ as was stated before. Then the
gradient flow (\ref{p33}) brings about
\begin{equation}
dP_{\alpha }/dt=[[D_{\alpha },P_{\alpha }],P_{\alpha }]-\sum\limits_{\beta
=1,\;\beta \neq \alpha }^{k}(P_{\alpha }D_{\alpha }P_{\beta }+P_{\beta
}D_{\alpha }P_{\alpha }).  \label{p35}
\end{equation}%
The flow (\ref{p35}) one can also obtain as a Hamiltonian one with respect
to the following canonical symplectic structure on $\mathcal{P}$:

\begin{equation}
\omega ^{(2)}(P)=\sum\limits_{\alpha =1}^{k}Sp(P_{\alpha }dP_{\alpha }\wedge
dP_{\alpha }P_{\alpha })  \label{p36}
\end{equation}%
with an element $P=\underset{\alpha =1}{\overset{k}{\oplus }}P_{\alpha }\in
\mathcal{P}$, natural constraints $P_{\alpha }P_{\beta }=P_{\alpha }\delta
_{\alpha \beta },$ $\alpha ,\beta =\overline{1,k},$ and some Hamiltonian
function $H\in D(P)$ which has to be found making use of the expression
\begin{equation}
-i_{\nabla _{\varphi }\Psi }\omega ^{(2)}=dH.  \label{p37}
\end{equation}%
Straightforward but tedious calculations of (\ref{p37}) give rise to the
same expression (\ref{p35}).

The system of equations (\ref{p35}) is a natural generalization of the
replicator dynamics for description of a multi-agent interaction. They can
describe for example economic communities trying to adapt to a changing
environment. In this approach agents update their behavior in order to get
maximum payoff under the given matrices of strategies $\{a_{ik}^{(\alpha
)}\} $ in response to the information received from other agents. The
structure of the system (\ref{p35}) is quite different from the system of
equations (\ref{p1a}). The first term on the right hand side describes the
individual evolution of each economic agent $\alpha \,$in accordance to its
own independent replicator dynamics and the second term describes average
payoffs of all others $k$ agents except $\alpha $.

\end{document}